# Irreversibility and Measurement in Quantum Mechanics

DOUGLAS M. SNYDER
LOS ANGELES, CALIFORNIA

In quantum mechanics, in principle a human observer does not require a macroscopic physical instrument to make a measurement. A good deal of evidence supporting this thesis comes from work on macroscopic quantum tunneling (Das Sarma, Kawamura, & Washburn, 1995). It is often maintained that there is nothing unique that a human observer can do that a physical, macroscopic measuring instrument, often of simple design, cannot also accomplish. It is generally argued that when a macroscopic measuring instrument is used, irreversibility characterizes the measurement process. The measurement cannot be reversed even if a human observer does not immediately see or know the result obtained with the measuring instrument.[1]

This irreversibility has been ascribed to an increase in entropy that occurs when a measurement is made (e.g., van Hove, 1959). It is implied that irreversibility is a common element in quantum mechanical measurement, even characterizing the circumstances where a person might make a measurement unaided by a macroscopic measuring device.

In a related manner, Bohr (1935) maintained that quantum mechanical measurement also depended on the interaction between a macroscopic measuring instrument and the physical existent measured. He noted that when a macroscopic physical measuring apparatus is used, there is inevitably some loss of information concerning the measured system due to the resulting physical interaction. Bohr presented a gedankenexperiment involving a particle passing through a slit in a diaphragm that is part of an experimental apparatus. In one scenario in which the position of the particle passing through the diaphragm is precisely determined, knowledge of the particle's momentum is lost because of the rigid connection between the diaphragm and the rest of the apparatus necessary to establish the position of the diaphragm when the particle passes through. In another scenario, in which the momentum of the particle is precisely determined, knowledge of the position of the particle is lost because of the flexible connection between the diaphragm and rest of the apparatus necessary to determine the momentum of the diaphragm before and after the

---

[1] Generally, this irreversibility means that it is highly unlikely that the physical interaction that is the measurement could occur in the opposite direction of time to the one in which it is occurring or has occurred.

# Irreversibility and Measurement

particle passes through the diaphragm. For Bohr, once the information is lost in the measurement process, the measurement cannot be reversed.

It is known that the consideration of a physical system as a macroscopic system instead of as a collection of microscopic quantum mechanical systems is arbitrary. If the physical system is considered in the latter manner, the Schrödinger equation provides the basis for a lawful delineation of the physical existent that was to be measured by the macroscopic apparatus and that microscopic part of the physical system designated the measurement apparatus with which the system that was to be measured interacts. The lawful manner in which these microscopic physical systems function, including their interaction, is reversible. Information is not lost, and the concept of entropy, dependent on a macroscopic physical system, is not relevant.

Thus when physicists maintain that irreversibility in measurement is a sufficient condition to ensure that a measurement has indeed occurred, the following feature of quantum mechanical measurement is not addressed. Whether a physical system is considered: (1) macroscopic in nature (so as to be available to measure a microscopic physical system) or (2) a collection of microscopic physical systems some portion of which interact with the microscopic system of interest in a reversible fashion in accordance with the Schrödinger equation, is an arbitrary decision on the part of the individual considering the overall experimental circumstances. That is, the presence of irreversibility depends on whether a measurement is being made, and whether or not a measurement is being made depends on an arbitrary decision by the experimenter as how the experimental circumstances are structured.

It should be noted that the assumption that irreversibility characterizes a human observer who does not require a macroscopic physical instrument to make a quantum mechanical measurement is unwarranted in the absence of specifying the neurophysiological mechanism, presumably functioning like a macroscopic physical apparatus, by which an irreversible measurement is made. Also, given the flexibility in considering a macroscopic physical system as such or as a combination of many microscopic systems, this result holds when a macroscopic measuring device is used in making a measurement as well. It would be important to specify how this neurophysiological mechanism involves those physical processes that would be responsible for a macroscopic physical instrument's ability to engage in an irreversible measurement. It would be these processes that would presumably underlie the neurophysiological mechanism.

# Irreversibility and Measurement

Another concern should be noted. How do irreversible measurement processes, presumably based on physical interaction, account for "negative" observations (Epstein, 1945; Renninger, 1960)? In a negative observation, a measurement occurs in the absence of a physical interaction between a measuring instrument and the existent measured, with an accompanying change in the wave function describing the existent measured.[2]

Because of the above considerations, one has to doubt that irreversibility in quantum mechanical measurement has been given an adequate foundation. Indeed, the flexibility in the measurement process as regards the observer's ability to consider a physical system as a macroscopic measuring instrument or instead as a large system comprised of many macroscopic systems indicates the importance of a subjective element.

## REFERENCES

Bohr, N. (1935). Can quantum-mechanical description of nature be considered complete? *Physical Review*, *49*, 1804-1807.

Cook, R. J. (1990). Quantum jumps. In E. Wolf (Ed.), *Progress in Optics* (Vol. 28) (pp. 361-416). Amsterdam: North-Holland.

Das Sarma, S., Kawamura, T., and Washburn, S. (1995). Resource letter QIMA-1: Quantum interference in macroscopic samples. *American Journal of Physics*, *8*, 683-694.

Epstein P. (1945). The reality problem in quantum mechanics. *American Journal of Physics*, *13*, 127-136.

Renninger, M. (1960). Messungen ohne Störung des Meßobjekts [Observations Without Changing the Object]. *Zeitschrift für Physik*, *158*, 417-421.

van Hove, L. (1959). The ergodic behaviour of quantum many-body systems. *Physica*, *25*, 268-276.

---

[2] A negative observation depends on the possibility of an alternative observation involving a physical interaction. Indeed, Cook (1990) noted the possibility of a negative observation in which one existent (a light photon) involved in the physical interaction can be registered by the unaided human eye.